# Online Similarity-and-Independence-Aware Beamformer for Low-Latency Target Sound Extraction

Atsuo Hiroe, *Member, IEEE*

**Abstract** This study introduces an online target sound extraction (TSE) process using the similarity-and-independence-aware beamformer (SIBF) derived from an iterative batch algorithm. The study aimed to reduce latency while maintaining extraction accuracy. The SIBF, which is a linear method, provides more accurate estimates of the target than an approximate magnitude spectrogram reference. The transition to an online algorithm reduces latency but presents challenges. First, contrary to the conventional assumption, deriving the online algorithm may degrade accuracy as compared to the batch algorithm using a sliding window. Second, conventional post-processing methods intended for scaling the estimated target may widen the accuracy gap between the two algorithms. This study adopts an approach that addresses these challenges and minimizes the accuracy gap during post-processing. It proposes a novel scaling method based on the single-channel Wiener filter (SWF-based scaling). To further improve accuracy, the study introduces a modified version of the time-frequency-varying variance generalized Gaussian distribution as a source model to represent the joint probability between the target and reference. Experimental results using the CHiME-3 dataset demonstrate several key findings: 1) SWF-based scaling effectively eliminates the gap between the two algorithms and improves accuracy. 2) The new source model achieves optimal accuracy, corresponding to the Laplacian model. 3) Our online SIBF outperforms conventional linear TSE methods, including independent vector extraction and minimum mean square error beamforming. These findings can contribute to the fields of beamforming and blind source separation.

*Index Terms*— **target sound extraction, similarity-and-independence-aware beamformer, online algorithm, low latency, single-channel Wiener filter**

## I. INTRODUCTION

Target sound extraction (TSE) involves estimating a target (sound of interest) from one or more mixtures of multiple sound sources. It significantly contributes to improving human speech intelligibility and supporting automatic speech recognition (ASR) [1]–[3]. TSE methods are generally classified as linear and nonlinear. Recent advancements in deep neural networks (DNNs) have significantly bolstered nonlinear methods. These DNN-based TSE methods can extract clean speech from overlapping utterances of multiple speakers or noisy sources using clues such as enrollment speech, video of a human face, and text queries [4]–[7]. Conversely, linear methods excel at avoiding nonlinear distortions, including musical noises and spectral distortions [3], [8], [9]. Linear methods can be further categorized into two: beamformers (BFs) that utilize time-frequency (TF) masks, known as mask-based BFs [10]–[12]; and TSE methods based on independent component analysis (ICA) [13], termed ICA-based TSE, including independent vector extraction (IVE) [14]–[16] and maximum likelihood distortionless response (MLDR) [17]–[19]. Mask-based BFs typically derive the estimated target without iteration, whereas ICA-based TSE achieves higher accuracy through iterative refinement [17], [18].

Adopting ICA-based TSE, we previously developed a novel method termed the *similarity-and-independence-aware beamformer* (SIBF) [20], [21]. The SIBF employs a magnitude spectrogram generated by any target-enhancing method, including DNN-based TSE, as a reference for the target. Considering that magnitude spectrograms can be derived from various data types, including waveforms and TF masks, the SIBF can integrate with different DNN outputs. This flexibility enables the SIBF to refine DNN outputs from diverse DNN sources. These usage scenarios require the SIBF to estimate the target more accurately than the reference, namely the DNN output. This method leverages both the dependence on the reference and the independence of other sources than the target to satisfy this requirement. This refinement effect is a unique advantage of the SIBF over other ICA-based methods, which do not conform to the above requirement.

Previously, we employed an iterative batch algorithm, known as *batch SIBF*, which processes all sound data within a recorded segment. However, in practical scenarios, this algorithm faces two major challenges.

1.  The latency in obtaining the SIBF output increases with

the duration of the target [22] because the extraction process cannot commence until the target recording is complete. Consequently, downstream tasks, including ASR, may experience latency exceeding the combined recording and SIBF processing times.

2. Only a single filter is estimated in the segment. Thus, the extraction accuracy tends to be lower if the sources are nonstationary or moving [23].

This study addresses the first issue, although the employed approaches aim to mitigate both challenges. Thus, the study aims to minimize latency while maintaining extraction accuracy, in contrast to batch SIBF.

A possible solution to both issues involves updating the extraction filter, which generates the estimated target from the observations, at short intervals (e.g., per frame). This approach requires updating the following processes per frame: 1) reference generation, 2) filter estimation, 3) pre-processing, and 4) post-processing.

For reference generation, the following approaches are applicable: the framework of the low-latency speech separation method using convolutional neural networks (CNNs) [24] and that using long short-term memories (LSTMs) [25]. Therefore, this study focuses on the remaining processes.

However, we found that deriving the variant for the per-frame update from the batch SIBF may cause issues in both the filter estimation and post-process. Both issues have not yet been investigated, although other ICA-based methods using the iterative batch algorithm face similar challenges.

One issue is the accuracy degradation caused by filter estimation. This occurs uniquely when the original batch algorithm is iterative, highlighting the inequivalence between two methodologies for updating the extraction filter per frame: 1) the batch algorithm using a sliding window, known as the *windowed batch algorithm* [26], and 2) the online algorithm [26], [27]. Conventionally, these two methodologies were assumed to be equivalent; thus, only the latter has been examined because it contains significantly smaller computational costs. However, this equivalence is not satisfied if the windowed batch algorithm is iterative. Therefore, deriving the online algorithm may cause accuracy degradation.

The other is expanding the accuracy gap between the two algorithms during post-processing, which adjusts the scale of the estimated target. The conventional scaling method based on the minimal distortion principle (MDP) [28] may degrade the estimated target generated by the online algorithm owing to its sensitivity to residual interferences present in the estimated target, although this aspect has not been thoroughly investigated. Given that other scaling methods do not conform to the SIBF framework, a novel method is required.

Considering these issues, we rephrase the goal of this study as follows: to reduce latency compared with the batch algorithm and maintain the extraction accuracy compared with not only the batch algorithm but also the windowed batch algorithm, which may outperform the online algorithm contrary to the conventional assumption. This study adopts an approach that minimizes the accuracy gap in post-processing while using the online algorithm to achieve this. Instead of employing MDP-based scaling, the study proposes a novel scaling method that avoids gap expansion and mitigates accuracy degradation.

Furthermore, this study examines a source model that characterizes the joint probability between the target and reference signals to improve extraction accuracy. Specifically, we introduce the time-frequency-varying variance (TV) generalized Gaussian model [29]–[31], which has demonstrated superior performance in blind source separation (BSS) than that of the original TV Gaussian model. Building on insights from our previous study utilizing TV Gaussian, TV Student's t, and bivariate spherical (BS) Laplacian models, we modify the TV generalized Gaussian model accordingly.

The remainder of this paper is organized as follows: Section II explains the formulation of the SIBF. Section III reviews the techniques employed in the online TSE methods to derive the SIBF updated per frame. Section IV presents the aforementioned issues and their solutions. Section V explains the implementation of the SIBF updated per frame, including filter estimation and the proposed scaling method. Section VI presents a sequence of experiments. Section VII discusses the experimental results. Finally, Section VIII presents the conclusions of this study.

## II. FORMULATION OF SIBF

Before exploring the derivation of the online SIBF, this section provides an updated formalization of the batch SIBF. Initially formulated in [20] as an extension of the deflationary ICA [13], the method was subsequently unified with the formulation of the mask-based BFs in [21]. This section first explains the unified formulation and subsequently examines source models while introducing the TV generalized Gaussian model.

The notations listed in Table I are consistently used throughout this paper to represent the TF domain signals, with $f$, $t$, and $k$ denoting the indices of the frequency bin, frame, and channel, respectively. These indices begin with 1.

TABLE I
SIGNAL NOTATIONS (VARIABLES $f$, $t$, AND $k$ DENOTE THE INDICES OF THE FREQUENCY BIN, FRAME, AND CHANNEL, RESPECTIVELY; AND $F$, $T$, $M$, AND $N$ DENOTE TOTAL NUMBERS OF FREQUENCY BINS, FRAMES, SOURCES, AND MICROPHONES, RESPECTIVELY).

| Signal name | Spectrogram | An element | Column vector of all channel elements |
|---|---|---|---|
| **Source** | $\boldsymbol{S}_k \in \mathbb{C}^{F \times T}$ | $s_k(f,t) \in \mathbb{C}$ | $\boldsymbol{s}(f,t) = [s_1(f,t),\ldots,s_M(f,t)]^{\mathrm{T}}$ |
| **Observation** | $\boldsymbol{X}_k \in \mathbb{C}^{F \times T}$ | $x_k(f,t) \in \mathbb{C}$ | $\boldsymbol{x}(f,t) = [x_1(f,t),\ldots,x_N(f,t)]^{\mathrm{T}}$ |
| **Estimated target** | $\boldsymbol{Y} \in \mathbb{C}^{F \times T}$ | $y(f,t) \in \mathbb{C}$ | (Not available) |
| **Reference** | $\boldsymbol{R} \in \mathbb{R}^{F \times T}$ | $r(f,t) \in \mathbb{R}$ | (Not available) |

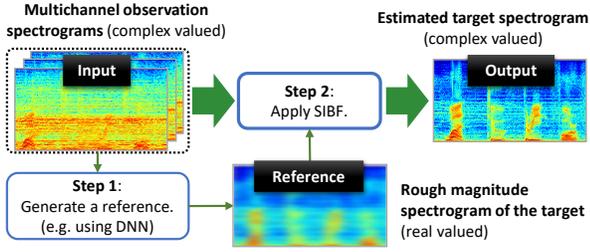

Fig. 1 Workflow of SIBF

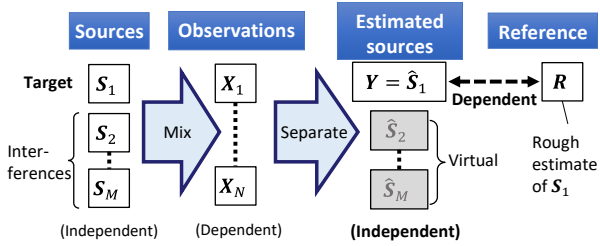

Fig. 2 Framework of SIBF

Each spectrogram comprises $F$ frequency bins and $T$ frames. The estimated target, after adjusting its scale, is referred to as the *SIBF output* to distinguish it from the estimated target $Y$. Moreover, to emphasize the assignment to a variable, the left arrow '←' is used instead of '=' as needed.

Fig. 1 illustrates the workflow of the SIBF. The inputs are the multichannel observation spectrograms obtained from multiple microphones, while the output is a spectrogram of the estimated target. A magnitude spectrogram associated with the target is used as the reference and can be estimated using various methods, including the DNN-based TSE. The workflow involves two steps: 1) estimating the rough magnitude spectrogram of the target, and 2) applying the SIBF with the rough spectrogram as the reference. The first step is independent of the SIBF framework explained below.

### A. Framework of SIBF

Fig. 2 illustrates the framework of the SIBF, including the signal mixing and separation processes. The framework assumes that $M$ sources $S_1, \ldots, S_M$ are mutually independent. Without loss of generality, $S_1$ is considered the target source, namely the source of interest in this study, whereas the other sources are considered interferences. The observations $X_1, \ldots, X_N$ represent the spectrograms obtained from $N$ microphones. In the TF domain, each observation spectrogram $X_k$ is approximated as an instantaneous mixture of sources. From the assumption of the sources, the estimated sources $\hat{S}_1, \ldots, \hat{S}_M$ are also mutually independent. Considering that only $\hat{S}_1$ is of interest, this is represented as $Y$. To certainly associate $Y$ with the target source $S_1$, the dependence between $Y$ and the reference $R$, as well as the independence of all the estimated sources, are leveraged. The objectives are to make $Y$ both similar to the reference using the dependence and more accurate than the reference using the independence. Deflationary estimation, that is, one-by-one separation, is employed to generate only $Y$. This indicates that the other estimated sources, $\hat{S}_2, \ldots, \hat{S}_M$, are virtual. In each frequency bin, $Y$ is estimated as follows.

$$y(f, t) = w(f)^{\mathrm{H}} x(f, t), \tag{1}$$

where $w(f) \in \mathbb{C}^{N \times 1}$ is a filter that generates $y(f, t)$ from $x(f, t)$. Herein, $w(f)$ is referred to as the *extraction filter*.

### B. Estimating Extraction Filter

The TSE problem, shown in Fig. 2, can be solved using maximum-likelihood (ML) estimation [32]. In our previous study [21], we elucidated that the SIBF formulation is equivalent to the following constrained minimization problem:

$$w(f) = \arg\min_{w(f)} \left\{ -\sum_t \log \mathrm{P}\big(r(f, t), y(f, t)\big) \right\}, \tag{2}$$

$$\mathrm{s.\,t.}\, \frac{1}{T} \sum_t |y(f, t)|^2 = 1, \tag{3}$$

where $\mathrm{P}(\cdot)$ denotes a *source model* that is a joint probability density function between the given reference and target used to represent the dependence between them. The constraint given by (3) helps avoid the trivial solution of $w(f) = 0$.

We previously demonstrated that $w(f)$ can be estimated as the eigenvector corresponding to the minimum eigenvalue in the following generalized eigenvalue (GEV) problem and that the differences in the source models appear only in computing the weight $c(f, t)$.

$$\Phi_x(f) = \frac{1}{T} \sum_t x(f, t) x(f, t)^{\mathrm{H}}, \tag{4}$$

$$\Phi_c(f) = \frac{1}{T} \sum_t c(f, t) x(f, t) x(f, t)^{\mathrm{H}}, \tag{5}$$

$$w(f) = \mathrm{GEV}_{\min}(\Phi_c(f), \Phi_x(f)), \tag{6}$$

where $\Phi_x(f)$ and $\Phi_c(f)$ are the observation and weighted covariance matrices, respectively, and $\mathrm{GEV}_{\min}(A, B)$ denotes the eigenvector $v$ corresponding to the minimum eigenvalue $\lambda_{\min}$ in (7).

$$Av = \lambda_{\min} Bv. \tag{7}$$

Whether estimating $w(f)$ requires iteration depends on the source model used.

### C. Source Models

Since the extraction performance depends on the source model used, selecting a proper model is significant. Here, we overview existing source models used in the SIBF framework and present a modified version of the TV generalized Gaussian model.

#### 1) Overviewing Existing Source Models

We previously examined the TV Gaussian, TV Student's t, and BS Laplacian models as candidates for the source models, represented as (8), (9), and (10), respectively.

$$\mathrm{P}\big(r(f,t), y(f,t)\big) \propto \exp\left(-\frac{|y(f,t)|^2}{\max\left(r(f,t)^\beta, \varepsilon\right)}\right), \quad (8)$$

$$\mathrm{P}\big(r(f,t), y(f,t)\big) \propto \frac{1}{r(f,t)^2}\left(1 + \frac{2}{\nu}\frac{|y(f,t)|^2}{r(f,t)^2}\right)^{-\frac{2+\nu}{2}}, \quad (9)$$

$$\mathrm{P}\big(r(f,t), y(f,t)\big) \propto \exp\left(-\sqrt{\alpha r(f,t)^2 + |y(f,t)|^2}\right), \quad (10)$$

where $\max(\cdot)$ denotes selecting the maximum argument; $\beta$, $\nu$, and $\alpha$ are hyperparameters to control the influence of the reference, referred to as the *reference exponent*, *degree of freedom*, and *reference weight*, respectively; and $\varepsilon$, referred to as a *clipping threshold*, is a small positive value to avoid division by zero.

Our previous study reported several significant findings for these models. For the TV Gaussian model, we faced difficulties in finding the optimal $\beta$; two accuracy peaks appeared in both $0 < \beta < 2$ and $\beta > 2$, whereas the case $\beta = 2$ exhibited the lowest, although only this setting strictly corresponds to this model. We discussed that one peak unexpectedly resulted because varying $\beta$ changed the possibility of the clipping in (8), not identifying which peak corresponded to this. However, this model contained the advantage that $\boldsymbol{w}(f)$ can be obtained without iterations, which facilitates tuning hyperparameters common over all source models. Therefore, this study tackles the dual peak issue.

Meanwhile, the TV Student's t and BS Laplacian models both outperformed the TV Gaussian model in extraction accuracy owing to the iteration. Nonetheless, we could not determine which model was better; the BS Laplacian model outperformed the other if it was optimally tuned, although the optimal $\alpha$ within (10) sensitively depended on the noisiness of the scenario. Therefore, this study explores a different source model that outperforms the two models.

*2) Modified Version of TV Generalized Gaussian Model*

In the BSS field, the TV generalized Gaussian model has been employed to achieve better separation accuracy [29]–[31]. In this study, this model is applied to the SIBF with several modifications reflecting on the findings of the TV Gaussian model mentioned in Section II.C.1) and compared with the existing three models.

The modified version of the TV generalized Gaussian model can be described as follows:

$$\mathrm{P}\big(r(f,t), y(f,t)\big) \propto \exp\left(-\left(\frac{|y(f,t)|}{r'(f,t)^\beta}\right)^\rho\right), \quad (11)$$

where $\rho > 0$, referred to as the *shape parameter*, determines the shape of the distribution, and $r'(f,t)$ denotes the clipped reference calculated in (12).

$$r'(f,t) = \max(r(f,t), \varepsilon). \quad (12)$$

Compared with the original formula presented in [29] and [31], (11) is unique in the following aspects:

1. The clipping mechanism is included similar to (8), given that the reference $r(f,t)$ can be zero in the SIBF framework, and the clipping threshold $\varepsilon$ within (12) is treated as a significant hyperparameter that affects extraction accuracy.
2. Contrary to (8), the reference is clipped before being powered to prevent the dual peak issue mentioned in Section II.C.1).

Here, the formulas for estimating $\boldsymbol{w}(f)$ are derived. The derivation step is significant for this study because it facilitates understanding the dependency changes between the batch and online algorithms mentioned in Section IV.A. For simplicity, this study focuses on the case of $0 < \rho \le 2$. The cases of $\rho = 2$ and 1 are called the TV Gaussian and Laplacian models, respectively. In the TV Gaussian case, $\boldsymbol{w}(f)$ can be obtained without iterations using (4), (13), (5), and (6).

$$c(f,t) = \frac{1}{r'(f,t)^{2\beta}}. \quad (13)$$

Conversely, when $\rho < 2$, the extraction filter lacks a closed-form solution. In this case, iterative rules may be formulated utilizing the auxiliary function algorithm [33], [34]. After substituting (11) into (2), the following inequality is obtained [29]:

$$\sum_t \left(\frac{|y(f,t)|}{r'(f,t)^\beta}\right)^\rho \le \sum_t \left\{\frac{\rho|y(f,t)|^2}{2b(f,t)^{2-\rho}r'(f,t)^{2\beta}} + \left(1 - \frac{\rho}{2}\right)b(f,t)^\rho\right\}, \quad (14)$$

where $b(f,t)$ denotes a positive value referred to as the *auxiliary variable*. Differing from the original inequality used in [29], (14) includes the term $r'(f,t)^\beta$ in the denominator. Instead of minimizing the left-hand side in (14), this algorithm minimizes the right-hand side alternatively for $b(f,t)$ and $\boldsymbol{w}(f)$. For $b(f,t)$, (1) and (15) are calculated for all $t$.

$$b(f,t) = \frac{|y(f,t)|}{r'(f,t)^\beta}. \quad (15)$$

For $\boldsymbol{w}(f)$, (16), (4), (5), and (6) are computed.

$$c(f,t) = \frac{1}{b(f,t)^{2-\rho}r'(f,t)^{2\beta}} \quad (16)$$

$$= \frac{1}{r'(f,t)^{\beta\rho}|y(f,t)|^{2-\rho}} \quad (17)$$

Consequently, $\boldsymbol{w}(f)$ can be obtained by iterating (1), (17), (5), and (6). In particular, $c(f,t)$ for the TV Laplacian model is described as follows:

$$c(f,t) = \frac{1}{r'(f,t)^\beta|y(f,t)|}. \quad (18)$$

We can start the iteration by estimating $\boldsymbol{w}(f)$ using the TV Gaussian model with the best hyperparameters. This methodology is referred to as the *boost start* [21].

### 3) Two Interpretations of TV Gaussian Model

We mention two possible interpretations of the variance regarding the TV Gaussian model. Each study that utilizes this model, including [17], [18], and [35], adopts one of the following two interpretations:

1. In [35] and within the SIBF framework, the variances are predetermined for each frame and considered constants. For instance, in this study, the variances correspond to the denominators in (8). This interpretation eliminates the need for an iterative process to determine $w(f)$.

2. In [17] and [18], both the variances and the extraction filter are variables to be estimated. Such an interpretation necessitates an iterative algorithm that updates both sets of parameters in an alternating manner.

This study treats the second interpretation as a particular case using source models other than the TV Gaussian. For example, the case of $\rho = 0$ makes (17) identical to (19), which regards $|y(f,t)|^2$ as an estimate of the variance.

$$c(f,t) = \frac{1}{|y(f,t)|^2}. \quad (19)$$

Additionally, we found that (19) can also be derived using the TV Student's t model with $\nu \to 0$.

## III. OVERVIEWING EXISTING LINEAR TSE METHODS

This section overviews methodologies for deriving algorithms used in existing linear TSE methods, such as mask-based BFs and ICA-based ones. Additionally, we observe methods that adjust the output scale of TSE methods, considering that this study challenges a conventional scaling method and proposes a new one. In this section, $c(f,t)$ represents any weight used in computing weighted covariance matrices.

### A. Mask-Based BFs

Online algorithms have been proposed for the following mask-based BFs:

1. Maximum signal-to-noise ratio (max-SNR) or GEV BF [27], [36], [37];
2. Minimum variance distortionless response (MVDR) BF [38]–[42];
3. Minimum mean square error (MMSE) or multichannel Wiener filter (MWF) BF [23], [43].

Regardless of the BF used, common methodologies are applied to derive the corresponding online algorithms as follows: The mask-based BFs use one or two weighted covariance matrices of the observations to estimate the extraction filter. Observing the algorithms that compute these matrices, we classify them into four categories, as illustrated in Fig. 3. We explain each algorithm in this figure to emphasize the difference between (b) and (c), which is the focus of this study.

Fig. 3 (a) illustrates the batch algorithm, which computes the weighted covariance matrices from all the data included in the segment, and solely computes the extraction filter from the matrices.

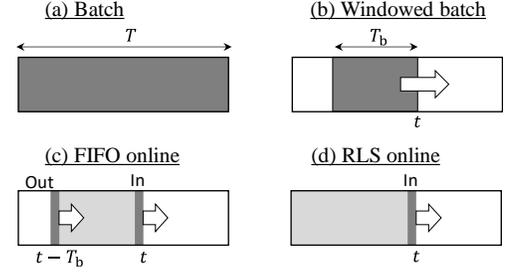

Fig. 3 Derivation steps from the batch algorithm to the online algorithm; (a) batch, (b) windowed batch, (c) FIFO online, and (d) RLS online algorithms

Next, the algorithms that compute covariance matrices and extraction filters per frame are explained. Let $w(f,t)$ and $\Phi(f,t)$ be an extraction filter and covariance matrix computed at frame $t$, respectively. Depending on the BF used, $\Phi(f,t)$ can represent the target, interference, or observation covariance matrix. The estimated target $y(f,t)$ is computed as follows:

$$y(f,t) = w(f,t)^{\mathrm{H}} x(f,t). \quad (20)$$

The filter $w(f,t)$ is estimated from $\Phi(f,t)$ depending on the BF used, and $\Phi(f,t)$ is computed depending on the algorithm used. Parts (b)–(d) in Fig. 3 illustrate three algorithms: the windowed batch [26], first-in-first-out (FIFO) online [26], [27], and recursive least square (RLS) online [36], [41], [44], which are represented as (21), (22), and (23), respectively.

$$\Phi(f,t) = \\ (1-g) \sum_{\tau=0}^{T_{\mathrm{b}}-1} g^{\tau} c(f,t-\tau) x(f,t-\tau) x(f,t-\tau)^{\mathrm{H}}, \quad (21)$$

$$\Phi(f,t) = g\Phi(f,t-1) \\ + (1-g)\{c(f,t)x(f,t)x(f,t)^{\mathrm{H}} \\ - g^{T_{\mathrm{b}}} c(f,t-T_{\mathrm{b}})x(f,t-T_{\mathrm{b}})x(f,t-T_{\mathrm{b}})^{\mathrm{H}}\}, \quad (22)$$

$$\Phi(f,t) = g\Phi(f,t-1) \\ + (1-g)c(f,t)x(f,t)x(f,t)^{\mathrm{H}}, \quad (23)$$

where $T_{\mathrm{b}}$ and $g$ denote the window length and a forgetting factor such that $0 < g < 1$, respectively. The weight $c(f,t)$ can be a mask value for the target or interferences, otherwise the case $c(f,t) = 1$ indicates computing an observation covariance matrix corresponding to (4).

The windowed batch algorithm computes $\Phi(f,t)$ in (21) by using a sliding window. The FIFO online algorithm, represented in (22), reduces the computational cost by leveraging $\Phi(f,t-1)$, although being equivalent to (21). Note that this equivalence holds only when $c(f,t)$ is independent of $y(f,t)$. The RLS online algorithm approximates the FIFO approach when $T_{\mathrm{b}}$ is sufficiently large. This has been extensively used in online BFs as it does not require buffering the past observations and weights.

TABLE II
CLASSIFICATION OF ICA-BASED TSE

| Method | Constraint | Batch | Online |
|--------|-----------|-------|--------|
| SIBF | Unit valiance | [20], [21] | This study |
| IVE | Dependence on all bins | [14]–[16] | [55], [57] |
| MLDR & WPD | Distortionless response | [17], [19], [61] | [18], [62] |

The following techniques are employed to further reduce the computational cost:

- In the MVDR and MMSE BFs, matrix inversion lemma (MIL) is applied to efficiently update the inverse of $\boldsymbol{\Phi}(f,t)$ [45], [46].
- In the max-SNR BF, an iterative method referred to as the *power method* (PM), is applied to solve the GEV problem efficiently [47], [48].

The online algorithms in this study comprise the FIFO and RLS online algorithm, excluding the windowed batch algorithm. The RLS online algorithm, combined with the PM and MIL, is treated as an approximation of the FIFO online algorithm to reduce the computational cost.

### B. ICA-Based TSE

This category includes IVE [14]–[16], MLDR [17]–[19], and the SIBF. Table II presents online algorithms proposed for both IVET and MLDR. We first examine IVE, discussing how it differs from SIBF, then provide an overview of MLDR.

#### 1) IVE

Considering that IVE originated from independent vector analysis (IVA) [49]–[52], we first explain IVA. The IVA framework was proposed to solve the frequency permutation problem, in which the order of the estimated sources is inconsistent among the frequency bins. This assumes that all the frequency bins included in the same spectrogram are mutually dependent. Whereas the gradient algorithm was initially employed [49], [50], the auxiliary function algorithm was applied to achieve faster and more stable separation [34], referred to as *AuxIVA*. Online variants of AuxIVA are proposed for real-time processing [22], [53].

IVE is an extension of IVA to the problem of extracting a non-Gaussian source. Analogous to the IVA, the study first employs the gradient algorithm [15], [54]. Subsequently, it applies the auxiliary function algorithm [55], [56] and proposes the online variants [55], [57].

In studies focusing on online IVA and IVE [22], [53], [55], [57], the online variants were directly derived by applying the RLS online methodology represented in (23). Thus, these investigations did not specifically address the accuracy degradation discussed in Section IV.A, which occurs between the windowed batch and FIFO online algorithms.

To extract only the target, IVE has further been extended to leverage a pilot signal associated with the target [55], [58]–[60], known as *piloted IVE*. The pilot signal is based on the identification of a dominant speaker within the mixture, including DNN-based embeddings known as *x-vectors* [55].

Although the reference $\boldsymbol{R}$ can be used for the piloted IVE, this combination differs from SIBF, as explained below. The piloted IVE framework leverages two constraints: 1) all frequency bins within $\boldsymbol{Y}$ are mutually dependent, and 2) the pilot is converted to a sequence of scalar values that depend on all the bins. If these constraints are represented using the SIBF formulation in (2), the piloted IVE can be formulated as (24), with the constraint represented in (3).

$$\boldsymbol{w}(f) = \arg\min_{\boldsymbol{w}(f)} \left\{ -\sum_t \log \mathrm{P}(\|\boldsymbol{R}(t)\|_2, \|\boldsymbol{Y}(t)\|_2) \right\}, \quad (24)$$

where $\boldsymbol{R}(t)$ and $\boldsymbol{Y}(t)$ denote the $t$th column vectors of $\boldsymbol{R}$ and $\boldsymbol{Y}$, respectively; $\|\cdot\|_2$ denotes the L2 norm of the given vector. Ignoring the reference type difference, the piloted IVE corresponds to a case in which the BS Laplacian model, represented as (10), is used in (24). We refer to TSE methods based on (24) as *IVE-constrained SIBF*, which is a generalization of the piloted IVE in terms of source models.

The weight $c(f,t)$ can be obtained by simply replacing $|y(f,t)|$ and $r(f,t)$ with $\|\boldsymbol{Y}(t)\|$ and $\|\boldsymbol{R}(t)\|$, respectively, in the formula for the SIBF (e.g., (18)). However, the L2 norm calculation presents the following challenges using this method: 1) the similarity between $\boldsymbol{Y}$ and $\boldsymbol{R}$ as a spectrogram is not guaranteed, and 2) the order between this calculation and temporal normalization, such as normalizing $r(f,t)$ and scaling $y(f,t)$ mentioned in Section V.B, can affect extraction accuracy.

#### 2) MLDR

The MLDR is an extension of the MVDR BF. It assumes that the target follows the TV Gaussian model to maximize the likelihood of the BF outputs under the distortionless response constraint. This category also includes studies on the weighted-power minimization distortionless response (WPD) [61] because WPD integrates MLDR and dereverberation [18]. Both MLDR and WPD use an iterative algorithm that alternatively estimates the extraction filter and variances. As mentioned in Section II.C.3), this study considers this algorithm as a specific case of the TV generalized Gaussian model. To estimate only the target, these methods employ a steering vector corresponding to the direction of the target.

The online variants of the MLDR and WPD were derived in [18] and [62], respectively. Similar to the online IVE, the RLS online algorithm was applied in the derivation. Thus, these studies did not investigate the accuracy degradation mentioned in Section IV.A.

### C. Scaling Methods

Several TSE methods require a post-process that adjusts the output scale because these methods cannot independently determine the scale. This subsection overviews existing scaling methods.

A scaling process is considered as multiplying the estimated target $y(f,t)$ by a scaling factor $\gamma(f)$:

$$y_{\text{scale}}(f,t) = \gamma(f)y(f,t), \tag{25}$$

where $y_{\text{scale}}(f,t)$ denotes the output of the TSE method after scaling. In the online manner, $\gamma(f)$ is replaced with $\gamma(f,t)$ to be updated per frame.

MDP-based scaling has extensively been used in the BSS field, including the IVA and IVE. This computes $\gamma(f)$ as follows [28]:

$$\gamma(f) = \frac{\sum_t x_m(f,t)\overline{y(f,t)}}{\sum_t |y(f,t)|^2}, \tag{26}$$

where $m$ and $\overline{y(f,t)}$ denote the reference microphone index and the conjugate of $y(f,t)$, respectively. Our previous study also employed this method because this applies to the case in which the number of output channels differs from that of microphones.

The projection-back (PB) method [63] is also used in the BSS field; however, directly applying it to SIBF is challenging because the PB needs to compute the inverse of the separation process for all the estimated sources although SIBF only estimates one, as illustrated in Fig. 2.

The blind analytical normalization [64] (BAN) is another scaling method. This is often employed in combination with the max-SNR BF, computing $\gamma(f)$ as follows:

$$\gamma(f) = \frac{\sqrt{\boldsymbol{w}(f)^{\text{H}}\boldsymbol{\Phi}_{\text{n}}(f)\boldsymbol{\Phi}_{\text{n}}(f)\boldsymbol{w}(f)/N}}{\boldsymbol{w}(f)^{\text{H}}\boldsymbol{\Phi}_{\text{n}}(f)\boldsymbol{w}(f)}, \tag{27}$$

where $\boldsymbol{\Phi}_{\text{n}}(f)$ denotes an estimated covariance matrix of the interferences. Note that BAN is hard to apply to the SIBF because $\boldsymbol{\Phi}_{\text{n}}(f)$ is not computed in the SIBF framework.

## IV. Issues and solution for deriving online algorithm for SIBF

The online variant of the SIBF can be derived by applying the FIFO and RLS online methodologies to the windowed batch algorithm, as mentioned in Section III.A. However, the derivation recalls two issues: 1) The derivation may degrade the extraction accuracy by changing data dependencies, and 2) the MDP-based scaling may increase the accuracy gap between the windowed batch and online algorithms. These findings are novelties in this study because conventional studies on the online ICA-based TSE have not investigated them. This section explains the two issues and proposes a solution.

### A. Issue 1: Accuracy Degradation by Changing Data Dependencies

As mentioned in Section III.A, the FIFO online algorithm is equivalent to the windowed batch algorithm only when $c(f,t)$ within (21) is independent of $y(f,t)$. In fact, the SIBF does not satisfy this condition except by using the TV Gaussian model. This changes the data dependencies as explained below.

Fig. 4 (a) shows the exceptional TV Gaussian case. The weighted covariance matrix $\boldsymbol{\Phi}_c(f,t)$ is computed using (21) for the windowed batch or (22) for the FIFO online. The weights $c(f,t-T_{\text{b}}+1)$ to $c(f,t)$ are computed from (13) and do not depend on $y(f,t)$ or $\boldsymbol{w}(f,t)$. Hence, the FIFO online algorithm is equivalent to the windowed batch algorithm; thus, $\boldsymbol{\Phi}_c(f,t)$ computed with (21) is identical to that computed with (22).

By contrast, other models exhibit distinct dependencies, as depicted in (b) and (c) in Fig. 4. Part (b) indicates the case of the windowed batch algorithm; the weights $c(f,t-T_{\text{b}}+1)$ to $c(f,t)$ are influenced by $\boldsymbol{w}(f,t)$ because $c(f,t-\tau)$ is calculated from both $y(f,t-\tau)$ and $r(f,t-\tau)$ depending on the source model used and $y(f,t-\tau)$ is computed as

$$y(f,t-\tau) = \boldsymbol{w}(f,t)^{\text{H}}\boldsymbol{x}(f,t). \tag{28}$$

Conversely, $\boldsymbol{w}(f,t)$ depends on the weights via $\boldsymbol{\Phi}_c(f,t)$. These interdependencies arise from the auxiliary function algorithm discussed in Section II. This algorithm performs alternating minimizations for $\boldsymbol{w}(f,t)$ and the auxiliary variables [33], [34], as mentioned in Section II.C.2.

However, the FIFO online algorithm represented in (22) changes the dependencies as illustrated in Fig. 4 (c); only $c(f,t)$ depends on $\boldsymbol{w}(f,t)$ because $\boldsymbol{\Phi}(f,t-1)$ is independent of $\boldsymbol{w}(f,t)$ in (22), whereas $c(f,t-T_{\text{b}}+1)$ to $c(f,t)$ affect $\boldsymbol{w}(f,t)$. In the context of the auxiliary function algorithm, it can be inferred that the minimization for the auxiliary variables is incomplete and that the problem of minimizing (2) is not solved. This fact may degrade extraction accuracy for the FIFO online algorithm compared with the windowed batch algorithm. The same issue applies to the RLS online algorithm because this approximates the FIFO approach.

This finding requires a methodology change for online TSE approaches. Conventionally, most studies on the online IVA and ICA-based TSE omitted evaluating the windowed batch algorithm [18], [22], [53], [55], [57], [62] because these studies assumed this algorithm should be equivalent to the FIFO online algorithm; thus, the RLS online algorithm can approximate both. However, the equivalence does not hold except for the TV Gaussian case. Therefore, the accuracy gap between the windowed batch and FIFO online algorithms should be a significant concern in the online TSE

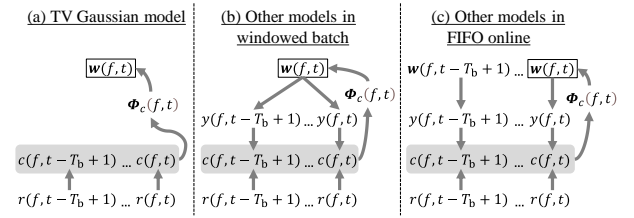

Fig. 4 Dependencies between the extraction filter $\boldsymbol{w}(f,t)$ and associated signals for (a) the TV Gaussian model in both windowed batch and FIFO online configurations; (b) other models examined within the windowed batch context; and (c) other models investigated using the FIFO online approach.

approach. Consequently, the four algorithms illustrated in Fig. 3 need to be evaluated.

Note that ICA-based TSE methods using the TV Gaussian may also cause degradation if the variance of this model is interpreted as a parameter to be estimated [17], [18] because this case corresponds to a particular case using other source models, as mentioned in Section II.C.3).

### B. Issue 2: Expanding Accuracy Gap Caused in MDP-Based Scaling

We also found that the MDP-based scaling may expand the accuracy gap between the windowed batch and online algorithms by further degrading the latter.

To consider possible issues with the MDP-based scaling, we conceptually decompose the signals described in (26); $x_m(f, t)$ and $y(f, t)$ are decomposed into two components as follows:

$$x_m(f, t) = x_{\text{tgt}}(f, t) + x_{\text{itf}}(f, t),\tag{29}$$

$$y(f, t) = y_{\text{tgt}}(f, t) + y_{\text{itf}}(f, t),\tag{30}$$

where $x_{\text{tgt}}(f, t)$ and $y_{\text{tgt}}(f, t)$ denote the components corresponding to the target, whereas $x_{\text{itf}}(f, t)$ and $y_{\text{itf}}(f, t)$ denote those corresponding to the other sources that are interferences. Herein, $y_{\text{itf}}(f, t)$ is considered the *residual interference*. From the independence assumption mentioned in Section II.A, the covariances between $x_{\text{tgt}}(f, t)$ and $y_{\text{itf}}(f, t)$, and between $y_{\text{tgt}}(f, t)$ and $x_{\text{itf}}(f, t)$ are considered to be 0. Thus, $\gamma(f)$ computed in (26) is conceptually identical to (31).

$$\gamma(f) = \frac{1}{T}\sum_t x_{\text{tgt}}(f, t)\overline{y_{\text{tgt}}(f, t)} + \frac{1}{T}\sum_t x_{\text{itf}}(f, t)\overline{y_{\text{itf}}(f, t)},\tag{31}$$

where the denominator in (26) has been replaced with $T$ because of (3).

The objective of the MDP-based scaling is to render the scale of $y(f, t)$ close to that of $x_{\text{tgt}}(f, t)$ using $x_m(f, t)$ [28]. If the extraction is perfectly executed, that is, $y_{\text{itf}}(f, t) = 0$, then (31) satisfies the objective. However, if $y_{\text{itf}}(f, t)$ persists, this formula tends to estimate an inaccurate scale owing to the second term in (31).

This issue can be particularly critical for the online algorithm owing to the accuracy degradation mentioned in Section IV.A. Essentially, this method may widen the accuracy gap between windowed batch and online algorithms because $y_{\text{itf}}(f, t)$ may persist more prominently in the latter algorithm. Therefore, a different scaling method is required to minimize this gap.

### C. Solution: SWF-Based Scaling as Alternative Post-process

An alternative scaling method must satisfy two aspects to address the two issues: 1) not expanding the accuracy gap, and 2) recovering the accuracy degradation by suppressing the interferences.

This study considers the scaling process as the formulation of the single-channel Wiener filter (SWF) [65], [66], formulated as follows:

$$\gamma(f) = \arg\min_{\gamma(f)}\sum_t |q(f, t) - \gamma(f)y(f, t)|^2\tag{32}$$

$$= \frac{1}{T}\sum_t q(f, t)\overline{y(f, t)},\tag{33}$$

where a complex number $q(f, t)$ denotes a reference signal for the scaling, known as the *scaling target*. Considering (1), we can also describe (33) as follows:

$$\boldsymbol{\varphi}_q(f) = \frac{1}{T}\sum_t q(f, t)\boldsymbol{x}(f, t)^{\text{H}},\tag{34}$$

$$\gamma(f) = \boldsymbol{\varphi}_q(f)\boldsymbol{w}(f),\tag{35}$$

where $\boldsymbol{\varphi}_q(f)$ denotes the covariance vector between $\boldsymbol{x}(f, t)$ and $q(f, t)$. Note that the MDP-based scaling is represented as a particular case in which $q(f, t) = x_m(f, t)$, although this is not proper as mentioned in Section IV.B.

The ideal candidate of $q(f, t)$ is $x_{\text{tgt}}(f, t)$ described in (29). This can conceptually be written as follows:

$$\gamma_{\text{ideal}}(f) = \frac{1}{T}\sum_t x_{\text{tgt}}(f, t)\overline{y_{\text{tgt}}(f, t)},\tag{36}$$

where $\gamma_{\text{ideal}}(f)$ denotes the ideal scaling factor, and scaling using $\gamma_{\text{ideal}}(f)$ is referred to as *ideal scaling*. This does not expand the accuracy gap because $\gamma_{\text{ideal}}(f)$ is not disturbed by $x_{\text{itf}}(f, t)$ or $y_{\text{itf}}(f, t)$ unlike (31). Thus, this resolves Issue 2 mentioned in Section IV.B. Considering that the SWF is effective in reducing noise [67]–[69], multiplying $y(f, t)$ by $\gamma_{\text{ideal}}(f)$ can eliminate the residual noise $y_{\text{itf}}(f, t)$. This can recover the accuracy degradation mentioned in Section IV.A as Issue 1. However, $x_{\text{tgt}}(f, t)$ is unknown in practical scenarios.

We propose using reference $r(f, t)$ combined with the phase of $x_m(f, t)$ as a scaling target:

$$q(f, t) = r(f, t)\frac{x_m(f, t)}{|x_m(f, t)|}.\tag{37}$$

If $q(f, t)$ remains significantly closer to $x_{\text{tgt}}(f, t)$ than $x_m(f, t)$, the SWF-based scaling can approximately act as the ideal scaling than the MDP-based scaling. Thus, this method can solve the two issues.

The novelty of this method lies in its approach. The MMSE BF combined with $q(f, t)$ implicitly includes this scaling mechanism because the formulation of this BF can be conceptually factorized into the MVDR BF and SWF-based scaling [68]–[70]. However, independently applying this scaling method to other linear TSE methods as a post-process is novel. Moreover, this method distinguishes itself from the SWF-based post-masking post-process [23], [66], [69], [70] represented in (38) and (39).

$$\mu_{\text{post}}(f, t) = \frac{|\hat{y}_{\text{tgt}}(f, t)|^2}{|\hat{y}_{\text{tgt}}(f, t)|^2 + |\hat{y}_{\text{itf}}(f, t)|^2}, \quad (38)$$

$$y_{\text{scale}}(f, t) = \mu_{\text{post}}(f, t) y(f, t), \quad (39)$$

where $\hat{y}_{\text{tgt}}(f, t)$ and $\hat{y}_{\text{itf}}(f, t)$ denote estimates of $y_{\text{tgt}}(f, t)$ and $y_{\text{itf}}(f, t)$, respectively. A post-mask $\mu_{\text{post}}(f, t)$ is non-negative and time-varying, whereas a scaling factor $\gamma(f)$ obtained with the batch algorithm is complex-valued and time-invariant. Essentially, combining SIBF with SWF-based scaling maintains linearity, enabling the advantages of linear methods discussed in Section I to be leveraged.

## V. Implementation

To explore the differences between the windowed batch and FIFO online algorithms mentioned in Section IV, we must implement and compare both alongside the batch and RLS online algorithms, even though this study treats the RLS approach as a candidate of the online SIBF. Given that the batch algorithm was explained in our previous study [21], this section explains the implementation of the remaining three algorithms, which update the extraction filter per frame. Additionally, we present the pre- and post-processing steps similarly, along with a pseudocode describing the operation of a TSE system using SIBF.

### A. Filter Estimation

The formulas for updating the extraction filter per frame are presented below. Let $\boldsymbol{\Phi}_c(f, t)$ and $\boldsymbol{\Phi}_x(f, t)$ denote the weighted and observation covariance matrices computed at frame $t$, respectively. The windowed batch, FIFO online, and RLS online algorithms compute $\boldsymbol{\Phi}_c(f, t)$ using (21), (22), and (23), respectively. Similarly, $\boldsymbol{\Phi}_x(f, t)$ is obtained using $c(f, t) = 1$ in each formula. Notably, (21) differs from (22) in computing $\boldsymbol{\Phi}_c(f, t)$, as mentioned in Section IV.A, whereas they are equivalent for $\boldsymbol{\Phi}_x(f, t)$. Using these matrices, the extraction filter $\boldsymbol{w}(f, t)$ can be estimated at frame $t$ as follows:

$$\boldsymbol{w}(f, t) = \text{GEV}_{\min}(\boldsymbol{\Phi}_c(f, t), \boldsymbol{\Phi}_x(f, t)). \quad (40)$$

In the mask-based BF field, several studies employing the RLS online algorithm report that the initial values of the co-variance matrices affect extraction accuracy [40]–[42]. To ensure stable estimation of initial values such as $\boldsymbol{\Phi}_c(f, 0)$ and $\boldsymbol{\Phi}_x(f, 0)$, this study computes them using the windowed batch algorithm, regardless of the algorithm used.

When using the RLS online algorithm, (40) can be computed more efficiently. We transform (40) into (41), given that the GEV problem corresponding to the minimum eigenvalue can be reformulated as a standard eigenvalue problem for the maximum eigenvalue [47].

$$\boldsymbol{\Phi}_c(f, t)^{-1}\boldsymbol{\Phi}_x(f, t)\boldsymbol{w}(f, t) = \frac{1}{\lambda_{\min}(f, t)}\boldsymbol{w}(f, t). \quad (41)$$

Applying the PM to (41) leads to the following update rules that iteratively obtain the eigenvector $\boldsymbol{w}(f, t)$:

$$\boldsymbol{w}(f, t) \leftarrow \boldsymbol{\Phi}_c(f, t)^{-1}\boldsymbol{\Phi}_x(f, t)\boldsymbol{w}(f, t), \quad (42)$$

$$\boldsymbol{w}(f, t) \leftarrow \frac{\boldsymbol{w}(f, t)}{\sqrt{\boldsymbol{w}(f, t)^{\mathrm{H}}\boldsymbol{\Phi}_x(f, t)\boldsymbol{w}(f, t)}} \quad (43)$$

Unlike the original PM [47], [48], (43) is executed to satisfy (3). The inverse of $\boldsymbol{\Phi}_c(f, t)$ within (41) can be computed by applying the MIL, which is written as follows [45], [46]:

$$(\boldsymbol{A} + d\boldsymbol{b}\boldsymbol{b}^{\mathrm{H}})^{-1} = \boldsymbol{A}^{-1} - \frac{\boldsymbol{A}^{-1}\boldsymbol{b}\boldsymbol{b}^{\mathrm{H}}\boldsymbol{A}^{-1}}{1/d + \boldsymbol{b}^{\mathrm{H}}\boldsymbol{A}^{-1}\boldsymbol{b}}, \quad (44)$$

where $\boldsymbol{A}$, $\boldsymbol{b}$, and $d$ denote an $N \times N$ full-rank matrix, $N \times 1$ vector, and a scalar value, respectively. If $\boldsymbol{\Phi}_c(f, t)$ is updated using (23), $\boldsymbol{\Phi}_c(f, t)^{-1}$ can be computed by the following assignments:

$$\boldsymbol{A} = g\boldsymbol{\Phi}_c(f, t-1), \quad (45)$$

$$\boldsymbol{A}^{-1} = \frac{1}{g}\boldsymbol{\Phi}_c(f, t-1)^{-1}, \quad (46)$$

$$\boldsymbol{b} = \boldsymbol{x}(f, t), \quad (47)$$

$$d = (1-g)c(f, t). \quad (48)$$

### B. Pre- and Post-Processes

In this study, both the pre- and post-processes are executed per frame. For both processes, the windowed batch algorithm is equivalent to the FIFO online algorithm, unlike the filter estimation.

The pre-process includes the reference normalization, which adjusts the square mean of the reference to 1 to render the optimal hyperparameters of source models independent of the scale of the reference. The normalized reference $r_{\text{norm}}(f, t)$ is obtained with (49) and employed in (12) instead of $r(f, t)$.

$$r_{\text{norm}}(f, t) = \frac{r(f, t)}{\sqrt{v_{\text{ref}}(f, t)}}, \quad (49)$$

where $v_{\text{ref}}(f, t)$ denotes the time mean of $r(f, t)^2$ computed depending on the algorithm used. For instance, the RLS online algorithm computes $v_{\text{ref}}(f, t)$ as follows:

$$v_{\text{ref}}(f, 0) = (1-g)\sum_{\tau=0}^{T_{\text{b}}-1} g^{\tau}r(f, -\tau)^2, \quad (50)$$

$$v_{\text{ref}}(f, t) = gv_{\text{ref}}(f, t-1) + (1-g)r(f, t)^2. \quad (51)$$

The post-process includes SWF-based scaling. When this process is executed per frame, it is described as follows, instead of (35) and (25):

$$\gamma(f,t) = \boldsymbol{\varphi}_q(f,t)^{\mathrm{H}} \boldsymbol{w}(f,t), \tag{52}$$

$$y_{\mathrm{scale}}(f,t) = \gamma(f,t) y(f,t), \tag{53}$$

where $\gamma(f,t)$ denotes the scaling factors in the $t$th frame; and $\boldsymbol{\varphi}_q(f,t)$ denotes the covariance vector between $\boldsymbol{x}(f,t)$ and $q(f,t)$ computed depending on the algorithm used. For instance, the RLS online algorithm computes $\boldsymbol{\varphi}_q(f,t)$ as

$$\boldsymbol{\varphi}_q(f,0) = (1-g) \sum_{\tau=0}^{T_{\mathrm{b}}-1} g^{\tau} \boldsymbol{x}(f,-\tau) \overline{q(f,-\tau)}, \tag{54}$$

$$\boldsymbol{\varphi}_q(f,t) = g \boldsymbol{\varphi}_q(f,t-1) \\ + (1-g) \boldsymbol{x}(f,t) \overline{q(f,t)}. \tag{55}$$

Additionally, the RLS online algorithm for the MDP-based scaling can be obtained by simply replacing $q(f,t)$ with $x_m(f,t)$ in (54) and (55).

### C. Pseudocode

Algorithm 1 presents the pseudocode that updates the extraction filter per frame. This encompasses three variations: 1) the windowed batch algorithm; 2) FIFO online algorithm; and 3) RLS online algorithms combined with the PM and MIL. Lines ending with '(*1)' are skipped when using the TV Gaussian model. The frequency index '$*$' indicates all the frequency bins. For example, Line 3 indicates that $\boldsymbol{x}(1,t)$ to $\boldsymbol{x}(F,t)$ are computed and buffered.

In Line 1, min $(\cdot)$ indicates selecting the minimum argument value. This line supports the case that the segment duration is smaller than the window length $T_{\mathrm{b}}$. Lines 2–5 prepare the observations and references when the frame index is zero or negative, as shown in (56) and (57).

$$\boldsymbol{x}(f,-\tau) = \boldsymbol{x}(f,T_{\mathrm{b}}-\tau), \tag{56}$$
$$r(f,-\tau) = r(f,T_{\mathrm{b}}-\tau), \tag{57}$$

where $0 \le \tau < T_{\mathrm{b}}$. Lines 6–9 compute the initial values by using the windowed batch algorithm, regardless of the algorithm used. Line 6 computes those for pre- and post-processes by using (50) and (54), respectively. Lines 7 and 9 compute the initial values of $\boldsymbol{\Phi}_x(f,t)$ and $\boldsymbol{\Phi}_c(f,t)$ by using (58) and (59), respectively.

$$\boldsymbol{\Phi}_x(f,0) = (1-g) \sum_{\tau=0}^{T_{\mathrm{b}}-1} g^{\tau} \boldsymbol{x}(f,-\tau) \boldsymbol{x}(f,-\tau)^{\mathrm{H}}, \tag{58}$$

$$\boldsymbol{\Phi}_c(f,0) \\ = (1-g) \sum_{\tau=0}^{T_{\mathrm{b}}-1} g^{\tau} c(f,-\tau) \boldsymbol{x}(f,-\tau) \boldsymbol{x}(f,-\tau)^{\mathrm{H}}. \tag{59}$$

| **Algorithm 1**: Per-frame SIBF output generation |
|---|
| 1:     $T_{\mathrm{b}} \leftarrow \min(T_{\mathrm{b}}, T)$ |
| 2:     **for** $t = 1$ to $T_{\mathrm{b}}$ **do** |
| 3:        Compute and buffer $\boldsymbol{x}(*,t)$. |
| 4:        Compute and buffer $r(*,t)$. |
| 5:     **end for** |
| 6:     Compute the initial values for the pre- and post-processes. |
| 7:     Compute $\boldsymbol{\Phi}_x(*,0)$. |
| 8:     Compute $\boldsymbol{w}(*,0)$ using the TV Gaussian model. |
| 9:     Compute $\boldsymbol{\Phi}_c(*,0)$ and its inverse. |
| 10:    **for** $t = 1$ to $T$ **do** |
| 11:       **if** $t > T_{\mathrm{b}}$ **then** |
| 12:          Compute (and buffer) $\boldsymbol{x}(*,t)$. |
| 13:          Compute (and buffer) $r(*,t)$. |
| 14:       **end if** |
| 15:       Normalize $r(*,t)$. |
| 16:       Update $\boldsymbol{\Phi}_x(*,t)$. |
| 17:       $\boldsymbol{w}(*,t) \leftarrow \boldsymbol{w}(*,t-1)$ (*1) |
| 18:       **for** $i = 1$ to $K_{\mathrm{aux}}$ **do** (*1) |
| 19:          Compute $y(*,t)$. (*1) |
| 20:          Compute $c(*,t)$ depending on the source model used. |
| 21:          Update $\boldsymbol{\Phi}_c(*,t)$. |
| 22:          Update $\boldsymbol{w}(*,t)$. |
| 23:       **end for** (*1) |
| 24:       Compute $y(*,t)$ using (20). |
| 25:       Adjust the scale of $y(*,t)$. |
| 26:    **end for** |
| (*1) This line is skipped when the TV Gaussian model is used. |

When source models other than the TV Gaussian model are used, Line 9 computes $y(f,-\tau)$ using (28) with $t = 0$ to obtain $c(f,-\tau)$, after Line 8 estimates $\boldsymbol{w}(t,0)$ using the TV Gaussian model.

The rest of the lines represent the process for each frame. In Lines 12 and 13, neither $\boldsymbol{x}(*,t)$ nor $r(*,t)$ require buffering in the RLS online algorithm. Line 15 executes the reference normalization explained in Section V.B. Line 16 updates $\boldsymbol{\Phi}_x(f,t)$ by using (21), (22), and (23), depending on the algorithm used. Line 17 is skipped in the TV Gaussian case, except when the PM is used. Lines 18–23 iterate the loop of $K_{\mathrm{aux}}$ iterations for estimating $\boldsymbol{w}(*,t)$ based on the auxiliary function algorithm. These lines are not iterated in the TV Gaussian case. Line 21 computes $\boldsymbol{\Phi}_c(*,t)$ depending on the algorithm employed. Line 22 computes $\boldsymbol{w}(*,t)$ using (40). If PM and MIL are used in the RLS online algorithm, the eigenvector in (40) is obtained iteratively using (42) and (43), while $\boldsymbol{\Phi}_c(f,t)^{-1}$ in (42) is updated with the MIL mentioned in Section V.A. Lines 24 and 25 generate the SIBF output in frame $t$. The scaling method in Line 25 is explained in Section V.B.

## VI. EXPERIMENTS

We conducted several experiments using the CHiME-3 dataset [1] to verify whether the online SIBF can reduce the latency while maintaining extraction accuracy. The experiments included the following steps:

1. Verifying accuracy degradation using MDP-based scaling (Section VI.B).
2. Investigating the effects of SFW-based scaling (Section VI.C).
3. Approximating the FIFO online algorithm by using the RLS online algorithm with the PM and MIL to further reduce latencies and computational costs (Sections VI.D and VI.E).
4. Comparing the results obtained from the CHiME-3 simulated test set with those of conventional methods (Section VI.F).

This study treats the RLS online algorithm as an approximation of the FIFO approach. Thus, the latter is verified to be comparable with the windowed batch algorithm in Section VI.C, then replaced with the former in Section VI.D. In this section, the window length $T_b$ is expressed in seconds as needed, and the RLS online algorithm includes the PM and MIL. We begin by detailing the common experimental setup and then present the results of each experiment sequentially.

### A. Experimental Setup

#### 1) Dataset

The CHiME-3 dataset comprises sound data recorded in four noisy environments using six microphones attached to a tablet device, along with clean speech recorded in a recording booth with the same device. Using this dataset corresponds to the case in which a TSE method was applied to a speech enhancement task, which can be interpreted to indicate that $S_1$ in Fig. 2 is the only speech source, whereas the others are non-speech sources [3]. The clean speech comprised 410 utterances from four speakers. This dataset is labeled as *dt05_bth*. Background (BG) noises were recorded in four noisy environments, such as a bus, a café, a pedestrian area, and a street junction.

We artificially mixed the clean speech with BG noise to prepare the observation data as a development set with various signal-to-noise ratios (SNRs). During the mixing, we applied four multipliers (0.25, 0.5, 1.0, and 2.0) to the noises to represent the four scenarios shown in Table III. For the observation data, the SNR scores also indicate the signal-to-distortion ratio (SDR) [71]. Each scenario comprises 1640 utterances.

In Section VI.F, we introduced another dataset called the *CHiME-3 simulated test set*, used for comparing the SIBF with other methods. This comprises 1320 utterances, which are a combination of 330 utterances from four speakers and four background noises. Table IV shows the number and duration of the utterances for each dataset.

#### 2) Experimental System

Fig. 5 illustrates the experimental system employed in this study, designed based on the CHiME-4 baseline system [2].

**TABLE III**
**FOUR SCENARIOS USED AS DEVELOPMENT SET**

| Scenario name | Relative noise level | Multiplier to BG | SNR (=SDR) [dB] |
|---|---|---|---|
| BG×0.25 | Least noisy | 0.25 | 14.05 |
| BG×0.5 | Less noisy | 0.5 | 8.03 |
| BG×1.0 | Noisier | 1.0 | 2.03 |
| BG×2.0 | Noisiest | 2.0 | -3.93 |

**TABLE IV**
**STATISTIC DATA FOR EACH DATASET (Dev.: Development, Std.: Standard deviation)**

| Data-set | Number of utterances | Shortest [s] | Longest [s] | Mean [s] | Std. [s] |
|---|---|---|---|---|---|
| Dev. | 1640 | 1.81 | 14.18 | 6.36 | 2.31 |
| Test | 1320 | 1.55 | 12.37 | 6.23 | 2.32 |

The bold and thin lines denote the flows of multichannel and single-channel data, respectively. The index pair $(f, t)$ is omitted for simpler notation. Multichannel observations in the time domain were converted to multichannel spectrograms by applying the short-time Fourier transform (STFT) with 1024 points (640 ms) and 256 shifts (160 ms). The reference was generated with the DNN, explained in Section VI.A.3), from $x_5(f, t)$. Given that the fifth microphone was closest to the speaker position, the setting of $m = 5$ was used in (37). This system can switch the scaling method between the MDP- and SWF-based methods.

The SIBF output $y_{scale}(f, t)$ was converted to the waveform by applying the inverse STFT (ISTFT) after replacing the frequency bin data under 62.5 Hz and over 7812.5 Hz with zeros. For evaluating the reference, $r(f, t)$ was converted to the waveform after being combined with the phase of $x_5(f, t)$. This also corresponds to the time-domain representation of the scaling target $q(f, t)$ calculated in (37).

We implemented the system in Python and utilized NumPy for matrix operations. The program was executed on a Linux PC with an Intel Core i7-7700K 4.2-GHz CPU and 32 GB RAM.

In this system, only the reference generation process using the DNN remained in the batch mode. Therefore, during experiments for the online algorithm, we first computed and stored the references for all the frames and then used the stored ones in Lines 4 and 13 of Algorithm 1.

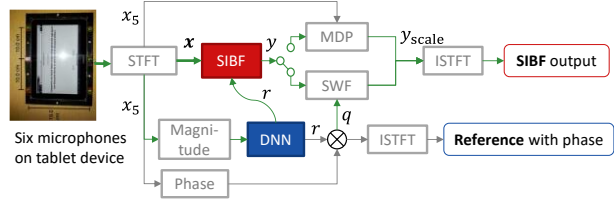

Fig. 5 Evaluation scheme for SIBF using CHiME-3 dataset

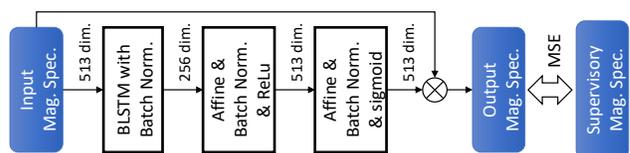

Fig. 6 DNN configuration with magnitude spectrogram (Mag. Spec.) output as reference. Numbers indicate input and output dimensions.

TABLE V
EXPERIMENTAL SETUP EXPLAINED IN DIFFERENT SUBSECTIONS (DEV.: DEVELOPMENT SET)

| Subsection | Algorithm | $T_b$ [s] | Source model | Scaling | Dataset | Metric |
|---|---|---|---|---|---|---|
| VI.B | Batch, windowed, FIFO | 5 (other than batch) | TV Gaussian, TV Laplacian | MDP | Dev. | $\Delta$SDR |
| VI.C | | | | SWF | | |
| VI.D | RLS+PM+MIL | 1–5 | TV Laplacian | | | |
| VI.E | All | 5 (other than batch), 2 (RLS only) | | | | RTF, latency |
| VI.F | Batch, RLS+PM+MIL | 2 (RLS only) | | | Test | PESQ, SDR, STOI, eSTOI |

TABLE VI
$\Delta$SDR USING MDP-BASED SCALING FOR BATCH, WINDOWED BATCH, AND FIFO ONLINE ALGORITHMS

| Source model | Algorithm | Iteration | $\Delta$SDR [dB] | | | | |
|---|---|---|---|---|---|---|---|
| | | | BG×0.25 (least noisy) | BG×0.5 (less noisy) | BG×1.0 (noisier) | BG×2.0 (noisiest) | Mean |
| TV Gaussian | Batch | - | 4.70 | 6.41 | 6.52 | 5.18 | 5.70 |
| | Windowed | - | 4.69 | 6.48 | 6.43 | 4.87 | 5.62 |
| | FIFO | - | 4.69 | 6.46 | 6.38 | 4.82 | 5.59 |
| TV Laplacian | Batch | 10 | 5.06 | 7.12 | 7.60 | 6.34 | 6.53 |
| | Windowed | 1 | 4.92 | 7.31 | 7.97 | 6.80 | 6.75 |
| | Windowed | 10 | 5.05 | **7.60** | **8.28** | **7.06** | **7.00** |
| | FIFO | 1 | **5.09** | 6.91 | 7.12 | 5.79 | 6.23 |
| | FIFO | 10 | **5.09** | 6.86 | 7.06 | 5.72 | 6.18 |
| Reference | - | - | 4.35 | 5.77 | 6.57 | 6.14 | 5.71 |

TABLE VII
$\Delta$SDR USING SWF-BASED SCALING FOR BATCH, WINDOWED BATCH, AND FIFO ONLINE ALGORITHMS

| Source model | Algorithm | Iteration | $\Delta$SDR [dB] | | | | |
|---|---|---|---|---|---|---|---|
| | | | BG×0.25 (least noisy) | BG×0.5 (less noisy) | BG×1.0 (noisier) | BG×2.0 (noisiest) | Mean |
| TV Gaussian | Batch | - | 6.08 | 8.97 | 10.39 | 9.71 | 8.79 |
| | Windowed | - | 6.08 | 9.22 | 10.61 | 9.67 | 8.89 |
| | FIFO | - | 6.10 | 9.23 | 10.60 | 9.66 | 8.90 |
| TV Laplacian | Batch | 10 | 6.12 | 9.13 | 10.80 | 10.50 | 9.14 |
| | Windowed | 1 | 5.79 | 9.06 | 10.88 | 10.75 | 9.12 |
| | Windowed | 10 | 5.84 | 9.26 | **11.12** | **10.94** | 9.29 |
| | FIFO | 1 | **6.40** | **9.40** | 10.97 | 10.46 | **9.31** |

### 3) DNN Configuration

Fig. 6 illustrates the DNN configuration used for reference generation. This was identical to that used in our previous study [21]. Both the input and output of the DNN consisted of magnitude spectrograms. The network was trained to extract a singular speech source from a composite of clean speech and background noise. Owing to the inclusion of a bidirectional LSTM layer within the network architecture, the inference process operated in batch mode.

### B. Verifying Accuracy Degradation Using MDP-Based Scaling

First, we verified the degree of accuracy degradation between the windowed batch and FIFO online algorithms mentioned in Section IV, using the setup described in the first row of Table V. MDP-based scaling was commonly used for the batch, windowed batch, and FIFO online algorithms. The

SDR was used as a metric that represents the extraction accuracy. We denote the difference between the SDR of the estimated target and that of the observation by $\Delta$SDR.

Before the evaluation, we conducted preliminary experiments for source model selection and model hyperparameter tuning, as mentioned in Appendices A and B. Consequently, we adopted the TV Laplacian model as the representative that achieved the best scores despite requiring an iterative process, as well as the TV Gaussian model, which obtains the extraction filter without iteration. The window length $T_b$ and forgetting factor $g$ were determined by employing the FIFO online algorithm with the TV Gaussian model; $T_b = 5$ s (312 frames) achieved the best $\Delta$SDR for $1 s \leq T_b \leq 10$ s, and $g = 0.99$ was the best among $0.95, 0.98$, and $0.99$.

Table VI shows the $\Delta$SDR scores for each source model and algorithm. The scores with 10 iterations were also described to evaluate the iterative improvement of the TV Laplacian model. The TV Laplacian model demonstrated different trends compared with the TV Gaussian model, as follows:

1. The windowed batch algorithm achieved the best scores except for the BG×0.25 scenario, whereas the three algorithms demonstrated almost the same scores for the TV Gaussian model.
2. Larger score differences between the windowed batch and FIFO online algorithms were observed as the scenario became noisier.
3. The iteration improved the scores for the windowed batch algorithm but slightly decreased the scores for the FIFO online algorithm.

Considering the equivalence between the windowed batch and FIFO online algorithms for the TV Gaussian case, we attribute the score differences between the second and third rows in Table VI to calculation errors occurring between (21) and (22).

The bottom row in Table VI shows the $\Delta$SDR scores of the reference in each scenario. The TV Gaussian model underperformed the reference in the BG×1.0 and BG×2.0 scenarios. Combining the TV Laplacian model with the FIFO online algorithm also underperformed the reference in the BG×2.0 scenario owing to the accuracy degradation. This row also represents the quality of the scaling target used for the SWF-based scaling, as mentioned in Section VI.A.2), and suggests that the scaling target was closer to the target source (clean speech) than the observation. Its significance is discussed in Section VI.B.

Here, we present several specific results of these experiments in Fig. 7 and Fig. 8, both comprising four signals, each of which is approximately 5 s long and represents both a waveform and a spectrogram. Each spectrogram is a close-up of the frequency bins between 100 and 1000 Hz. Parts (a) to (d) in Fig. 7 denote the target, mixture of interferences, observation, and reference, respectively. The target is the clean speech labeled as *F01_22GC010A* and includes two silent periods at both edges of the segment, as highlighted on the waveform. The mixture of interferences is background noise recorded in a bus and includes louder noise components in the frequency bins under 200 Hz, as highlighted on the spectrogram. We refer to them as lower-frequency noise. The observation is a mixture of the target and interferences and belongs to the BG×2.0 scenario in Table III. This also includes the lower-frequency noise; hence, the silent periods appear to be unclear. This signal is also used as the scaling target for MDP-based scaling. In (d), the lower-frequency noise is removed, and the silent periods at both edges can be identified, as highlighted on the waveform and spectrogram. This signal is used as the scaling target for SWF-based scaling.

Parts (a), (b), and (c) in Fig. 8 illustrate the SIBF outputs using MDP-based scaling, whereas (d) is explained in the next subsection. Parts (a) and (b) denote the results using the windowed batch algorithm with one and 10 iterations, respectively. We can observe the iterative improvement in the end-area of the segment; the lower-frequency noise is suppressed in (b), and the silent period can be identified, whereas the noise remains in (a). Part (c) denotes the results using the FIFO online algorithm with one iteration;

compared with (a) and (b), the lower-frequency noise remains more loudly and the silent periods on both edges are unclear. The result with 10 iterations is omitted because it appears to be identical to that with one iteration.

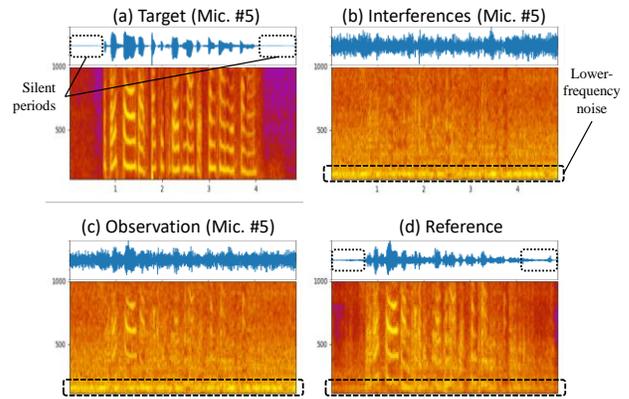

Fig. 7 Waveforms and spectrograms: (a) target, (b) mixture of interferences, (c) observation, and (d) reference

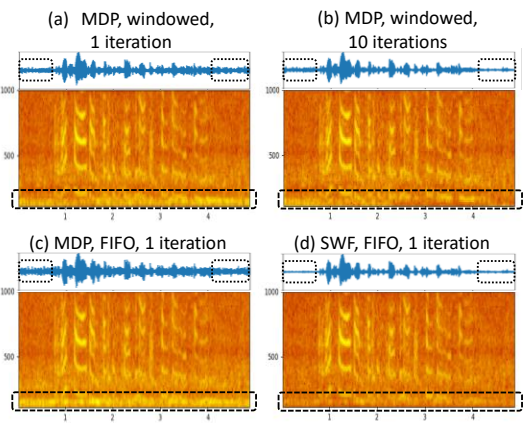

Fig. 8 Waveforms and spectrograms of the SIBF outputs: (a) using the MDP-based scaling and windowed batch with one iteration; (b) using the MDP-based scaling and windowed batch with 10 iterations; (c) using the MDP-based scaling and FIFO online with one iteration; and (d) using the SWF-based scaling and FIFO online with one iteration

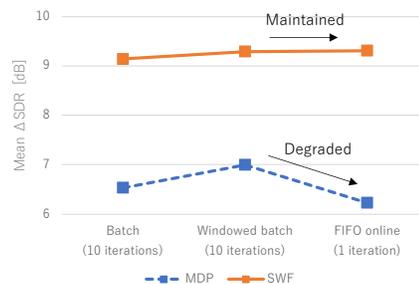

Fig. 9 Comparing MDP- and SWF-based scaling methods using TV Laplacian model; SWF-based scaling can prevent accuracy degradation between windowed batch and FIFO online algorithms and improve $\Delta$SDR scores for all algorithms.

| | ΔSDR [dB] | | | | |
|---|---|---|---|---|---|
| $T_b$ [s] | BG×0.25 (least noisy) | BG×0.5 (less noisy) | BG×1.0 (noisier) | BG×2.0 (noisiest) | **Mean** |
| 5 | **6.39** | **9.38** | **10.95** | 10.44 | **9.29** |
| 4 | 6.35 | 9.37 | **10.95** | 10.45 | 9.28 |
| 3 | 6.34 | 9.35 | 10.93 | **10.48** | 9.27 |
| 2 | 6.30 | 9.27 | 10.82 | 10.47 | 9.22 |
| 1 | 5.68 | 8.82 | 10.48 | 10.29 | 8.82 |

## C. Effects of SWF-based Scaling

Next, we confirmed whether the SWF-based scaling could eliminate the accuracy degradation using the setup described in the second row of Table V.

Table VII shows the ΔSDR scores for each source model and algorithm. The FIFO online algorithm with 10 iterations was excluded because the previous experiment did not demonstrate any improvement for this setup. Compared with Table VI, we can confirm that the SWF-based scaling improved the scores for all setups in Table VII and that this method reduced the accuracy gap between the windowed batch and the FIFO online algorithms for the TV Laplacian model. Comparing the windowed batch algorithm with 10 iterations and the FIFO online algorithm with one iteration, the maximum gap was 1.27 dB for the MDP-based scaling but 0.48 dB for the SWF-based scaling. Nevertheless, the FIFO online algorithm outperformed the windowed batch algorithm in the BG×0.25 and BG×0.5 scenarios. Consequently, observing the mean ΔSDR scores, the accuracy degradation for the TV Laplacian model was eliminated, unlike the MDP-based scaling case, as illustrated in Fig. 9. Therefore, we adopted SWF-based scaling in the subsequent experiments.

Here, we explain the effects of the SWF-based scaling in Fig. 8 (d), which illustrates the SIBF outputs of the FIFO online algorithm with one iteration. The results using the windowed batch algorithm with one and 10 iterations are omitted because these appear to be identical to (d). Unlike the MDP scaling case, the lower-frequency noise is suppressed throughout the segment, and the silent periods can be identified on both edges.

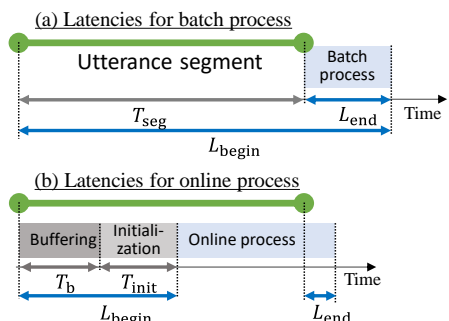

Fig. 10 Definitions of beginning-side latency ($L_{begin}$) and end-side latency ($L_{end}$)

## D. Approximating FIFO Online Using RLS Online

Next, we approximated the FIFO online algorithm using the RLS online algorithm combined with the PM and MIL mentioned in Section III.A. We also shortened the window length $T_b$ maintaining the extraction accuracy, considering that the length is directly linked to the latency.

The experimental setup is described in the third row of Table V. For the PM, we adopted two iterations because ΔSDR converged to the value almost the same as that using the GEV when the iteration count was greater than 1.

Table VIII shows the relationship between the ΔSDR scores and $T_b$ such that $1 \leq T_b \leq 5$ s (62 to 312 frames). The scores were almost constant except that $T_b = 1$ s. Comparing this table and Table VII suggests that the RLS online algorithm can approximate not solely the FIFO online algorithm but also the windowed batch algorithm, thanks to the SWF-based scaling.

Consequently, we adopted $T_b = 2$ s (125 frames) as the shortest length that can maintain the extraction accuracy.

## E. Measuring Latencies

We measured the processing time for all the algorithms using the setup described in the fourth row of Table V to verify whether the RLS online algorithm can reduce latencies. Note that the measured time did not include the reference generation time as mentioned in Section VI.A.2) and that the obtained latencies were estimated values.

Fig. 10 illustrates the two types of latencies:

1. Beginning-side latency $L_{begin}$, which is the duration from the beginning of the segment to the time when the first frame of the SIBF output is generated.
2. End-side latency $L_{end}$, which is the duration from the segment end to the time when the final frame of the SIBF output is generated.

For a system using the batch algorithm, $L_{end}$ is identical to the duration of the extraction process and $L_{begin}$ is the sum of $L_{end}$ and segment duration $T_{seg}$. Therefore, the latencies increase as the segment becomes longer. To simplify the estimation, the processing time of the batch algorithm for each segment was assumed to remain constant, irrespective of the $T_{seg}$. By contrast, for a system that generates the SIBF outputs per frame, both latencies are defined as follows:

$$L_{begin} = \min(T_b, T_{seg}) + T_{init},  \quad (60)$$

$$L_{end} = \max(L_{begin} - (1 - F_{rtf})T_{seg}, 0),  \quad (61)$$

where $T_{init}$ and $F_{rtf}$ denote the duration for the initialization process corresponding to Lines 6–9 in Algorithm 1, and real-time factor (RTF), that is, the mean ratio between the processing time and segment duration, respectively.

Table IX shows the estimated latencies for all algorithms, including the mean and worst-case (maximum) values. We verified that if $T_b = 2$ s, combining the RLS online algorithm with the PM and MIL can achieve fewer latencies than the batch algorithm in all cases. Additionally, the windowed

| Algorithm | Iteration | $T_b$ [s] | $T_{init}$ [s] | RTF | Mean [s] | | Worst [s] | |
|---|---|---|---|---|---|---|---|---|
| | | | | | $L_{begin}$ | $L_{end}$ | $L_{begin}$ | $L_{end}$ |
| Batch | 10 | - | - | 0.058 | 6.72 | 0.37 | 14.54 | 0.37 |
| Windowed | 1 | 5 | 0.11 | 6.613 | 4.71 | 47.05 | 5.11 | 98.75 |
| | 10 | 5 | 0.10 | 32.879 | 4.71 | 214.05 | 5.10 | 471.09 |
| FIFO | 1 | 5 | 0.10 | 0.815 | 4.70 | 3.83 | 5.10 | 4.67 |
| RLS+PM+MIL | 1 | 5 | 0.17 | 0.072 | 4.77 | 0.49 | 5.17 | 3.32 |
| | 1 | 2 | 0.08 | 0.072 | **2.08** | **0.00** | **2.08** | **0.32** |

batch algorithm contains a considerably larger computational cost, as its RTF is much greater than 1 in Table IX. Thus, using this algorithm in scenarios requiring low latencies is challenging despite its higher accuracies.

Combining the results from Sections VI.C and VI.E, we can infer that the online SIBF has not only reduced the latencies but also maintained the extraction accuracy compared with the batch and windowed batch algorithms.

### F. Comparison with Other Methods

In our final experiment, we compared the SIBF with several TSE methods, using the setup described in the bottom row of Table V. The experiment comprised the following two steps:

1. Comparing TSE methods that use a magnitude spectrogram as the reference, such as the SIBF, IVE-constrained SIBF mentioned in Section III.B.1), and MMSE BF, utilizing the batch algorithm.
2. Evaluating the three methods utilizing the RLS online algorithms and comparing them with other online TSE methods, such as MVDR and MLDR BFs.

This experiment used the CHiME-3 simulated test set and the following performance metrics: the SDR, narrowband perceptual evaluation of speech quality (PESQ) [72], short-time objective intelligibility measure (STOI) [73], and extended STOI (eSTOI) [74]. These metrics share the characteristic that a signal approximating clean speech yields a higher score.

#### 1) SIBF vs. IVE-Constrained SIBF vs. MMSE BF Using Batch Algorithm

First, we compared the batch SIBF, IVE-constrained SIBF, and MMSE BF, each of which can employ an approximately estimated magnitude spectrogram as the reference.

We commonly employed the TV Laplacian model and SWF-based scaling to compare the SIBF and IVE-constrained SIBF under unified conditions. Considering that the L2 norm calculation is affected by the scaling and reference normalization as mentioned in Section III.B.1), we calculated the weight $c(f, t)$ for IVE-constrained SIBF using (62)–(66).

$$\|\boldsymbol{R}(f)\|_2 = \sqrt{\textstyle\sum_f r(f, t)^2} \tag{62}$$

$$R_{norm}(t) = \frac{\|\boldsymbol{R}(f)\|_2}{\sqrt{v_R(t)}}, \tag{63}$$

$$R'(t) = \max(R_{norm}(t), \varepsilon), \tag{64}$$

$$Y'(t) = \sqrt{\textstyle\sum_f |y_{scale}(f, t)|^2}, \tag{65}$$

$$c(f, t) = \frac{1}{R'(t)^\beta Y'(t)}, \tag{66}$$

where $r(f, t)$, $v_R(t)$, and $y_{scale}(f, t)$ denote the reference without normalization, time mean of $\|\boldsymbol{R}(f)\|_2{}^2$ computed depending on the algorithm used, and output after applying the SWF-based scaling, respectively; $\varepsilon = 10^{-9}$ and $\beta = 1/4$, similar to the SIBF. We found that the above calculation achieved better scores than the case that used the normalized $r(f, t)$ and $y(f, t)$ in (62) and (65), respectively.

For the MMSE BF, we leveraged the scaling target $q(f, t)$ calculated in (37) as its reference. The extraction filter $\boldsymbol{w}_{mmse}(f)$ is obtained as follows [23], [75]:

$$\boldsymbol{w}_{mmse}(f) = \underset{\boldsymbol{w}(f)}{\arg\min} \sum_t \left| q(f, t) - \boldsymbol{w}(f)^H \boldsymbol{x}(f, t) \right|^2, \tag{67}$$

$$= \boldsymbol{\Phi}_x(f)^{-1} \boldsymbol{\varphi}_q(f), \tag{68}$$

where $\boldsymbol{\varphi}_q(f)$ is computed in (34). This BF includes the SWF-based scaling mechanism as mentioned in Section III.C.

The scores of the three are presented in Table X, in addition to the observation and reference. The top two rows correspond to the observation and reference. The observation SDR is substantively identical to the SNR. The third, fourth, and fifth rows show the scores of the SIBF, IVE-constrained SIBF, and MMSE BF, respectively. The SIBF outperformed the other two in all metrics.

#### 2) Online SIBF vs. Other Online TSE Methods

Next, we evaluated the three methods using the RLS online. The online SIBF was identical to that used in Section VI.E. The online variant of the IVE-constrained SIBF can be derived by combining (23) and (59) with (66). The online variant of the MMSE BF can be described as follows:

$$\boldsymbol{w}_{mmse}(f, t) = \boldsymbol{\Phi}_x(f, t)^{-1} \boldsymbol{\varphi}_q(f, t), \tag{69}$$

where $\boldsymbol{\varphi}_q(f, t)$ is computed in (54) and (55). The MIL was applied to compute $\boldsymbol{\Phi}_x(f, t)^{-1}$ in (69). We employed $g = 0.99$ and $T_b = 2\,\text{s}$ (125 frames), which were the same settings as those used in the online SIBF.

The sixth, seventh, and eighth rows in Table X correspond to the online SIBF, IVE-constrained SIBF, and MMSE BF,

TABLE X
PESQ, SDR, STOI, AND ESTOI SCORES USING CHIME-3 SIMULATED TEST SET (KF: KALMAN FILTER, GG: GENERALIZED GAUSSIAN)

| Method | Algorithm | Source model | Scaling | SDR [dB] | PESQ | STOI [%] | eSTOI [%] |
|--------|-----------|--------------|---------|----------|------|----------|-----------|
| Observation of Microphone #5 | - | - | - | 7.54 | 2.18 | 87.03 | 68.32 |
| Reference | Batch | - | - | 13.61 | 2.61 | 91.50 | 78.11 |
| Batch SIBF (this study) | Batch | TV Laplacian | SWF | 17.98 | 2.74 | **96.11** | **88.46** |
| Batch IVE-constrained SIBF | Batch | TV Laplacian | SWF | 17.65 | 2.72 | 95.62 | 87.21 |
| Batch MMSE BF | Batch | - | (SWF) | 14.54 | 2.54 | 93.75 | 81.96 |
| Online SIBF (this study) | RLS+PM+MIL | TV Laplacian | SWF | **18.09** | **2.75** | 96.03 | 88.30 |
| Online IVE-constrained SIBF | RLS+PM+MIL | TV Laplacian | SWF | 17.54 | 2.71 | 94.97 | 85.47 |
| Online MMSE BF | RLS+MIL | - | (SWF) | 14.41 | 2.54 | 93.48 | 81.36 |
| Cho+ 2021 [18] (Online MLDR BF) | RLS+PM+MIL | TV GG with $\rho = 0$ | Post-masking | - | 2.67 | - | 85.5 |
| Martín-Doñas+ 2020 [42] (Online MVDR BF) | RLS | - | KF | 14.89 | - | - | 84.1 |

respectively. The online SIBF outperformed the other two and was comparable with the batch SIBF.

The bottom two rows indicate the scores reported in studies on conventional TSE methods such as the online MLDR BF [18] and the online MVDR BF [42]. Note that these studies do not present all scores. Similar to the online SIBF, these methods update the extraction filter per frame. We regard the source model used in [18] as the TV generalized Gaussian model with $\rho = 0$, as mentioned in Section II.C.3. In [42], the MVDR BF was combined with the scaling method based on the Kalman filter (KF). The online SIBF outperformed these conventional online methods.

## VII. DISCUSSION

This section discusses the following aspects sequentially:

1. Behavior difference for each algorithm when using the MDP-based scaling
2. Behaviors of the SWF-based scaling
3. Findings on the TV generalized Gaussian model
4. Superiority of the SIBF to the IVE and MMSE BF

During the discussion, we denote the attributes of having a small or large absolute value as simply *small* or *large*.

### A. Behavior Difference for Each Algorithm When Using MDP-based Scaling

As mentioned in VI.B, the SIBF using the TV Laplacian model demonstrates largely different behaviors compared with the TV Gaussian model. This finding is a novel contribution of this study, and the same phenomena can be reproduced when using any models that require iterative batch algorithms. We discuss the reasons subsequently.

Comparing the batch and windowed batch algorithms in Table VI convinces us that the latter achieves slightly lower scores in the TV Gaussian case and better scores in the TV Laplacian case. We can explain this difference by discussing the iterative improvement unique to the TV Laplacian case. For the TV Laplacian model, the windowed batch algorithm practically increases the iteration counts compared with the original batch algorithm and effectively extracts the target more accurately. This insight is supported by the following: 1) in Table VI, the scores of this algorithm improve as the iteration counts increase, and 2) in Fig. 8 (b), the lower-frequency noise is suppressed in the end edge whereas remains in the beginning edge.

Meanwhile, comparing the windowed batch and FIFO online algorithms in Table VI convinces us that the latter achieves almost the same scores in the TV Gaussian case because of the equivalence between both algorithms, whereas lower scores in the TV Laplacian case. We can attribute this difference to the dependency change that uniquely occurs for the TV Laplacian model, as mentioned in Section IV.A. Additionally, this change deprives the iterative improvement, as verified in Section VI.B. This is because, in Fig. 4 (c), the updated $\boldsymbol{w}(f, t)$ hardly affects $\boldsymbol{\Phi}_c(f, t)$, given that only $c(f, t)$ is modified by $\boldsymbol{w}(f, t)$ unlike Fig. 4 (b). Therefore, the iteration cannot solve the accuracy degradation caused by deriving the FIFO online algorithm.

Furthermore, the accuracy gap due to the filter estimation is expanded by the MDP-based scaling, as mentioned in IV.B. The gap expansion can be confirmed in Fig. 8 (c), which is the result of the FIFO online algorithm; the lower-frequency noise remains loud compared with parts (a) and (b), which are the results of the windowed batch algorithm. We consider that this is pronounced in noisy frequency bins that satisfy $|x_{\text{tgt}}(f, t)| \ll |x_{\text{itf}}(f, t)|$ in (29), and that the residual interferences $y_{\text{itf}}(f, t)$ described in (30) remain substantial. For MDP-based scaling, the scaling factor within these bins is essentially influenced by the second term of (31) and is thus estimated to be significantly larger than the ideal value denoted in (36). Consequently, this method emphasizes residual interferences by applying a larger scaling factor, causing lower-frequency noise to persist.

### B. Behaviors of SWF-Based Scaling

The experimental results on the SWF-based scaling demonstrated the following trends:

1. The SWF-based scaling outperforms the MDP-based method.
2. This scaling method can eliminate the accuracy gap for the TV Laplacian model.

We discuss that both trends are caused because the SWF-based scaling can estimate the scaling factor more appropriately. As mentioned in VI.B, the scaling target represented in (37) outperformed the observation. Therefore, the scaling target for SWF-based scaling more closely approximates that of the ideal scaling compared with the MDP-based scaling. Consequently, this method is more akin to ideal scaling, as elaborated in Section IV.C. In the noisy frequency bins such that $|x_{\text{tgt}}(f,t)| \ll |x_{\text{itf}}(f,t)|$ in (29), the SWF-based scaling estimates a smaller factor that rarely emphasizes the residual interferences. We can confirm the improvement in Fig. 8 (d); the lower-frequency noise is hardly observed.

This discussion can also account for the second trend. Applying a small factor to $y(f,t)$ in (53) indicates that the effect of the extraction filter $w(f,t)$ is canceled, given that $y(f,t)$ is generated in (20). Therefore, the SWF-based scaling can nullify the accuracy degradation in the noisy bins. Considering that the SWF contains the effects of reducing the residual interference [67]–[69], the FIFO online algorithm can outperform the windowed batch algorithm, depending on the intensity of the reduction.

### C. Findings on TV Generalized Gaussian Model

This study introduced the TV generalized Gaussian model and made modifications reflecting the findings obtained in our previous study, as mentioned in Section II.C. Here, we discuss the behavior of this model, particularly focusing on the TV Gaussian and Laplacian cases, which correspond to the cases of $\rho = 2$ and $\rho = 1$ in (11), respectively.

For the TV Gaussian model, the dual peak issue mentioned in Section II.C.1) has been resolved in this study by clipping the reference before powering this. Appendix A indicates that the setting of $\beta = 1/4$ in (11) stably achieved peak accuracy when the clipping threshold $\varepsilon \le 10^{-4}$ in (12). We consider that these results are due to the balance of the following aspects:

1. A larger $\beta$ increases the sensitivity to $\varepsilon$ because this makes the characteristic of $c(f,t)$ in (13) closer to that of the binary mask thresholded with $\varepsilon$. This insight can explain the behavior for $\beta \ge 1/4$ in Fig. 11.
2. If $\beta$ is close to 0, $w(f)$ cannot be obtained properly because this case makes $\Phi_c(f)$ close to $\Phi_x(f)$ in (5) and any $w(f)$ can be a solution to (6). This analysis can explain the behavior for $\beta \le 1/4$ in Fig. 11.

Given that the TV Gaussian model has extensively been used in the fields of the BSS and ICA-based TSE, these findings can potentially contribute to both fields.

Meanwhile, the TV Laplacian model consistently outperformed the TV Gaussian model even when using the FIFO online algorithm, as mentioned in Sections VI.B and VI.C. Considering that deriving this algorithm deprives the iterative improvement, these results appear to be remarkable. We discuss the reason for this subsequently.

The SIBF using this model can approximately be interpreted as a variant of the TV gaussian case using an alternative reference. That is, (18) can be interpreted as using a combined value between $r(f,t)$ and $y(f,t)$, instead of the denominator of (13) as the reference in the TV Gaussian model. In this insight, if $y(f,t)$ is more accurate than $r(f,t)$, the combined value can work as a more accurate reference than $r(f,t)$ alone; thus, this model can outperform the TV Gaussian model even without the iterative improvement. However, whether this characteristic is maintained in any scenarios remains an open discussion.

### D. Superiority of SIBF to IVE and MMSE BF

Section VI.F suggests that the SIBF outperformed the IVE-constrained SIBF and MMSE BF in both the batch and online algorithms, although the three methods leverage the same reference. Considering that these methods commonly employ the SWF-based scaling, the differences in the extraction accuracy only arise from the filter estimation.

First, we discuss the IVE-constrained SIBF, which contains the following two characteristics similar to the piloted IVE, as mentioned in Section III.B.1):

1. Using $\|Y(t)\|_2$ imposes the constraint that all frequency bin data of the estimated target mutually depend.
2. Calculating $\|R(t)\|_2$ indicates that a single reference is shared among all frequency bins.

However, these characteristics rather disturb extraction when a magnitude spectrogram is available as a reference, because 1) the first characteristic can lead to accuracy degradation due to the model mismatch if the target does not satisfy this constraint [29], [76], and 2) calculating $\|R(t)\|_2$ reduces the reference quality, considering that a magnitude spectrogram comprises different data for each frequency bin.

By contrast, the SIBF does not impose the above constraint because leveraging the dependence between $r(f,t)$ and $y(f,t)$ can generate the outputs without the permutation ambiguity problem mentioned in Section III.B.1). Moreover, this method directly employs the magnitude spectrogram without reducing the quality.

Next, we discuss the MMSE BF. This method only approximates the reference in terms of the linear filtering represented in (67). Therefore, the scores for this method are similar to those for the reference, although this may appear to be superior or inferior to the reference in some metrics; from Table X, the MMSE BF appears to outperform the reference in the SDR, STOI, and eSTOI, while underperforming it in the PESQ.

Conversely, the SIBF framework has a characteristic that its output can outperform the reference by leveraging the independence of the sources, as mentioned in Section II.A. The experimental results support this characteristic.

## VIII. CONCLUSIONS

In this study, we detailed an online method for target sound extraction, referred to as online SIBF, to achieve both latency reduction and maintenance of extraction accuracy compared with batch SIBF. By examining techniques used in linear TSE methods, we derived an online algorithm from an existing iterative batch algorithm. However, this derivation introduced two challenges: 1) potential reduction in extraction accuracy due to changing dependencies between the extraction filter and related signals, which deviate from conventional assumptions; 2) increased potential for accuracy gap between the two algorithms in post-processing, as MDP-based scaling is sensitive to interferences present in both observations and the estimated target. These findings suggest that the windowed batch algorithm should remain a benchmark for achieving the highest extraction accuracy. They require methodology changes for online TSE approaches including the SIBF.

To address these issues and maintain optimal accuracy, we introduced a new scaling method, referred to as SWF-based scaling. This method uses a reference combined with the observation phase as the scaling target. As a result, this method more closely approximates ideal scaling than MDP-based scaling, thereby minimizing the accuracy gap between the windowed batch and online algorithms.

For improved extraction accuracy, we employed the TV generalized Gaussian model. Based on discussions of source models in our previous study, we modified this model to include both clipping and powering the reference.

To verify if this study can both reduce latency and maintain extraction accuracy, we conducted several experiments using the CHiME-3 dataset. We confirmed that the SWF-based scaling eliminated the accuracy gap between the two algorithms, which occurred prominently when using the MDP-based scaling. For the source model, the TV Laplacian model achieved the best accuracy as a variation of the TV generalized Gaussian model. Estimating the latencies for generating the SIBF output, the online SIBF reduced latencies on both the beginning and end sides. Therefore, by combining the online algorithm and SWF-based scaling, we could both reduce latencies and maintain extraction accuracy, compared with both the batch and windowed batch algorithms.

By comparing the online SIBF with other linear TSE methods, such as the MLDR, MVDR, and MMSE BFs, as well as the IVE using the reference, we confirmed that the SIBF outperformed others. Therefore, if a magnitude spectrogram is obtained as a reference of the target, the best combination is 1) using this for the SIBF filter estimation rather than the MMSE BF and IVE, and 2) reusing this as the scaling target for the SWF-based scaling.

Our future work includes the following aspects: 1) Building a system for real-time operation, with plans to implement online reference generation using DNNs. 2) Verifying the effectiveness of the online SIBF across various scenarios including a moving source case.

Finally, this study contributes to the BSS and BF fields in the following aspects:

1. The accuracy degradation caused by deriving the online algorithm from the iterative algorithm based on the auxiliary function, and the accompanying methodology change;

2. The sensitivity of the MDP-based scaling to the interferences included in both the observations and estimated target;

3. The behavior of the TV generalized Gaussian model when varying both the reference exponent and clipping threshold;

4. The superiority of the SIBF to the MMSE (or MWF) BF and IVE regarding extraction accuracy.

Overall, we anticipate that this study will stimulate additional research in both the BSS and BF fields.

## Appendix

This section includes tuning the hyperparameters of the TV generalized Gaussian model, which requires an iterative process to estimate the extraction filter. All experiments were conducted using the batch algorithm with MDP-based scaling, with the mean $\Delta$SDR across four scenarios used as an evaluation metric.

### A. Hyperparameter Tuning for TV Gaussian Model

Here, we detail the experiments on hyperparameter tuning for the TV Gaussian model, which is a specific model of the TV generalized Gaussian model with $\rho = 2$ in (11). Fig. 11 shows a plot of the mean $\Delta$SDR scores when both the reference exponent $\beta$ for $1/16 \leq \beta \leq 2$ and clipping threshold $\varepsilon$ for $10^{-10} \leq \varepsilon \leq 2$ are varied. The results appear to be classified into two types. For $\beta \geq 1/2$, the scores are sensitive to $\varepsilon$, while the peak score is almost the same. This type includes the conventional TV Gaussian model, which corresponds to the case of $\beta = 1$. Conversely, for $\beta \leq 1/4$, the scores were stable for $10^{-10} \leq \varepsilon \leq 10^{-5}$, although the highest score descends as $\beta$ becomes smaller. Given that the latter behavior is more robust to varied hyperparameters, we adopted the parameter set of $\beta = 1/4$ and $\varepsilon = 10^{-9}$ for being optimal one.

Significantly, the consistent demonstration of optimal scores when $\beta = 1/4$ for $\varepsilon \leq 10^{-4}$ indicates that the dual peak issue mentioned in Section II.C.1) has been resolved by clipping the reference before powering this in (12).

### B. Choosing Best Source Model Requiring Iterative Process

We detail the experiments of choosing the best source model. The experiments consisted of two steps: 1) hyperparameter tuning for the TV generalized Gaussian model, and 2) comparing the best one with the TV Student's t and BS Laplacian models.

First, we tuned the hyperparameters for the TV generalized Gaussian model with $\varepsilon = 10^{-9}$. The estimated targets were obtained with ten iterations, where the TV Gaussian model was used in the first iteration. Table XI displays the mean $\Delta$SDR scores for $\rho \in \{1/2, 1, 3/2\}$ and $\beta \in \{1/8, 1/4, 1/2, 1\}$, including those of the TV Gaussian model in the rightmost column. We identified the parameter, which set $\rho = 1$ and $\beta = 1/4$ as optimal. Additionally, the bottom-right cell in Table XI corresponds to the conventional TV Gaussian model, which yielded a score of 4.26 dB—significantly lower than the best score obtained for this model (5.70 dB).

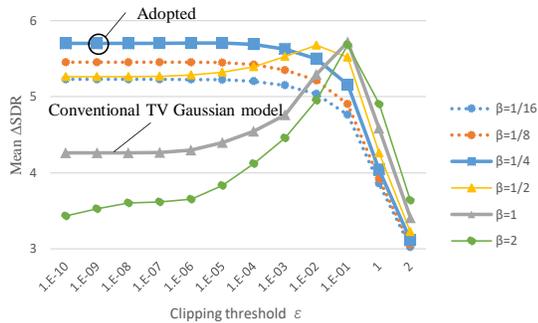

Fig. 11 Relationships between mean $\Delta$SDR and clipping threshold ($\varepsilon$) across various combinations of reference exponent values ($\beta$); Adopted parameters include $\beta = 1/4$ and $\varepsilon = 10^{-9}$.

TABLE XI

MEAN $\Delta$SDR [dB] FOR TV GENERALIZED GAUSSIAN MODEL OBTAINED USING DIFFERENT COMBINATIONS OF REFERENCE EXPONENT ($\beta$) AND SHAPE PARAMETER ($\rho$); ADOPTED PARAMETERS INCLUDE $\rho = 1$ AND $\beta = 1/4$.

| $\beta$ | $\rho$ | | | |
|---|---|---|---|---|
| | 1/2 | 1 (TV Laplacian) | 3/2 | 2 (TV Gaussian) |
| 1/8 | **6.45** | 6.46 | 6.12 | 5.45 |
| 1/4 | 6.44 | **6.53** | **6.30** | **5.70** |
| 1/2 | 6.34 | 6.19 | 5.77 | 5.26 |
| 1 | 5.98 | 5.14 | 4.51 | 4.26 |

TABLE XII

MEAN $\Delta$SDR FOR EACH SOURCE MODEL IN BATCH SIBF

| Source model | Hyperparameters | | | Mean $\Delta$SDR [dB] |
|---|---|---|---|---|
| | Iteration | $\rho$ | Others | |
| TV Gaussian | - | 2 | $\beta = 1/4$, | 5.70 |
| TV Laplacian | 10 | 1 | $\varepsilon = 10^{-9}$ | **6.53** |
| BS Laplacian | 10 | - | $\alpha = 100$ | 6.05 |
| TV Student's t | 10 | - | $\nu = 1$ | 6.35 |

Next, we compared the best scores for both models with those for the TV Student's t and BS Laplacian models. The former was evaluated with $\nu = 1$ in (9), and the latter with $\alpha = 100$ in (10). While these hyperparameters were identical to those used in our previous study [21], the experimental results slightly differed owing to modifications in the settings of the TV Gaussian model employed to estimate the initial extraction filter.

Table XII presents the mean $\Delta$SDR score for the four models. Both the TV Student's t and BS Laplacian models underperformed the TV Laplacian model while outperforming the TV Gaussian model. Therefore, we adopted the TV Laplacian model as the optimal choice, which requires an iterative process. We also chose the TV Gaussian model to rapidly tune the hyperparameters in common online processes across all source models.

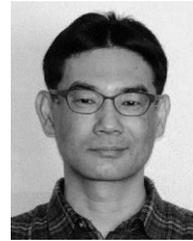

**Atsuo Hiroe** (M'11) received his B.S. and M.S. degrees in computer science from the Tokyo Institute of Technology (Tokyo, Japan) in 1994 and 1996, respectively.

From 1996 to 2014, he worked at Sony Corporation (Tokyo, Japan), conducting research and development on speech recognition, speech signal processing, and natural language understanding, among others. From 2014 to 2016, he was seconded to the National Institute of Information and Communications Technology (Kyoto, Japan) to study spoken dialog systems. Since 2016, he has been working at Sony Group Corporation, conducting research and development on speech signal processing.